\documentclass[aps,prl,superscriptaddress,twocolumn,10pt]{revtex4-2}

\usepackage{graphicx}% Include figure files
\usepackage{dcolumn}% Align table columns on decimal point
\usepackage{bm}% bold math
\usepackage{amsfonts}
\usepackage{tikz}
\usepackage[separate-uncertainty=true]{siunitx}
\usepackage{physics}
\usepackage{graphicx}
\usepackage{acronym}
\usepackage{xspace}
\usepackage{bbold}
\usepackage{braket}

\usepackage{datatool, filecontents}
\DTLsetseparator{,}% Set the separator between the columns.

\DTLloaddb[noheader, keys={thekey,thevalue}]{exportvalueerrors}{exportvalueerrors.dat}
\newcommand{\varr}[1]{\DTLfetch{exportvalueerrors}{thekey}{#1}{thevalue}}

\begin{document}
\begin{filecontents*}{exportvalueerrors.dat}
spinsplittingr,390

errspinsplittingr,4

GammaC2r,125

errGammaC2r,9

GammaC3r,101

errGammaC3r,18

psatC2r,5.8

errpsatC2r,1.2

psatC3r,3.4

errpsatC3r,1.6

etar,22.7

erretar,1.2

T1r,2.0

errT1r,0.7

maxfidelityr,0.806

errmaxfidelityr,0.004

maxinitrater,387

errmaxinitrater,10

spinsplitting,100

errspinsplitting,5

Afig3r,38.47

errAfig3r,0.12

Gamma20r,320

errGamma20r,120

Afig4ar,41.5

errAfig4ar,0.7

Apfig4ar,5.47

errApfig4ar,0.17

Gammafig4alr,3.2

errGammafig4alr,0.4

Gammafig4arr,4.3

errGammafig4arr,0.5

Gammafig4blr,2.0

errGammafig4blr,0.3

Gammafig4brr,2.6

errGammafig4brr,0.5

\end{filecontents*}

	
	\preprint{APS/123-QED}
	
	% \title{Strongly coupled $^{13}$C nuclear spin of silicon vacancy center inside a nanodiamond with sub-megahertz linewidth}
	\title{Strongly Coupled Spins of Silicon-Vacancy Centers Inside a Nanodiamond with Sub-Megahertz Linewidth}
	\author{M. Klotz}
	\altaffiliation{These authors contributed equally.}
	\affiliation{Institute for Quantum Optics, Ulm University, 89081 Ulm, Germany}
	\author{R. Waltrich}
	\altaffiliation{These authors contributed equally.}
	\affiliation{Institute for Quantum Optics, Ulm University, 89081 Ulm, Germany}
	\author{N. Lettner}
	\affiliation{Institute for Quantum Optics, Ulm University, 89081 Ulm, Germany}
	\affiliation{Center for Integrated Quantum Science and Technology (IQST), Ulm University,
	Albert-Einstein-Allee 11, 89081 Ulm, Germany.}
	\author{V. N. Agafonov}
	\affiliation{GREMAN, UMR 7347 CNRS, INSA-CVL, Tours University, 37200 Tours, France}
	\author{A. Kubanek}
	\email[Corresponding author: ]{alexander.kubanek@uni-ulm.de}
	\affiliation{Institute for Quantum Optics, Ulm University, 89081 Ulm, Germany}
	
	\date{\today}% It is always \today, today,

	\begin{abstract}		
		The search for long-lived quantum memories, which can be efficiently interfaced with flying qubits is longstanding. One possible solution is to use the electron spin of a color center in diamond to mediate interaction between a long-lived nuclear spin and a photon. Realizing this in a nanodiamond furthermore facilitates the integration into photonic devices and enables the realization of hybrid quantum systems with access to quantum memories. Here, we investigated the spin environment of negatively-charged Silicon-Vacancy centers in a nanodiamond and demonstrate strong coupling of its electron spin, while the electron spin's decoherence rate remained below \SI{1}{\mega\hertz}. We furthermore demonstrate multi-spin coupling with the potential to establish registers of quantum memories in nanodiamonds. 
		%		Order of magnitude recuded dephasing rate at liquid helium temperature compared to low-strain bulk \\
		%			--> optical/microwave coherent control possible \\			
		%			--> high sensitivity due to small linewidth?\\		
		%		Strongly coupled next-nearest neighbour C13 nuclear spin \\
		%		Coupled multi qubit system with moderate coherences \\
	\end{abstract}
	
	\maketitle
In the future, quantum based networks can provide secure communication or distributed quantum computing \citep{Kimble2008, Wehner2018, Briegel1998, Childress2006}. 
One of the remaining challenges is finding a scalable network node which can process, distribute and store quantum information, efficiently. Qubits based on solid-state quantum emitters offer advantages in terms of scalability. First small networks, for example based on negatively-charged Nitrogen-Vacancy centers in diamond (NV$^-$) have been realized in a pioneering work \citep{Pompili2021}. However, the NV$^-$ is prone to perturbations from external fields and the rate of coherent photons is low \citep{Faron2012, Chu2014}.
In contrast, group-IV defects like the negatively-charged Silicon Vacancy center (SiV$^-$) are insensitive to external electric fields and show intrinsically identical emitter \citep{Rogers2014, Jahnke2014, Waltrich2023}. Recent results demonstrated coherent control of the SiV$^-$ spin with coherence times in the ms range when operating at mK temperatures \citep{Sukachev2017}. Increasing the operation temperature is desirable to reduce technical overhead. A potential solution are SiV$^-$ in nanodiamonds (NDs) with modified electron-phonon interactions \citep{Klotz2022} which can further be integrated in hybrid quantum systems such as photonic crystal cavities \citep{Fehler2020, Fehler2021}. 

%In this letter we show the observation of strongly coupled spins in a ND, where one of them is in good agreement to theoretical modeling of a nearest neighbor $^{13}C$ nuclear spin \citep{nizovtsev2020}. 
In this letter, we show the observation of spins in a ND strongly coupled to the electron spin of a SiV- center. The coupling strength of one of the spins is in good agreement to theoretical modeling of a nearest neighbor $^{13}$C nuclear spin \citep{nizovtsev2020}.
We further show that for a SiV$^-$ in a ND the main mechanism for decoherence of its spin qubit, which is phonon-mediated dephasing, can already be mitigated at temperatures of around 4K. The resulting suppressed decoherence rate, access to a local memory and fast initialization rates lay the foundations for coherent control of an integratable hybrid quantum network node based on SiV$^-$s in NDs.

\begin{figure}[t]
		\centering	
		\includegraphics{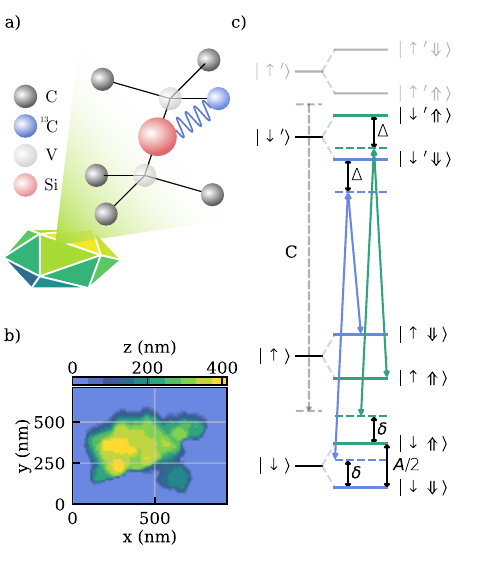}
		\caption{ a) Illustration of a SiV$^-$ with a coupled $^{13}$C nuclear spin inside a ND. The implied shear represents the presence of strain in the host crystal. b) AFM scan of the nanodiamond agglomeration containing the SiV$^-$s under consideration. c) Partial level scheme for transition C of the SiV$^-$. The four levels originating from the Zeeman effect are labeled as $\ket{\downarrow}$, $\ket{\uparrow}$, $\ket{\downarrow'}$, $\ket{\uparrow'}$. The corresponding hyperfine levels are indicated with an additional $\Downarrow$ or $\Uparrow$. Lambda systems connecting the correspondingly involved electron and nuclear spins are differently colored. Single- and two-photon detunings are labeled with $\Delta$ and $\delta$, respectively. Other transitions arising from A, B and D are not shown for simplicity.}
		\label{fig:Figure1}
	\end{figure}

The SiV$^-$ is a point defect in the diamond lattice, where a Silicon atom (Si) with an excess electron is situated between two adjacent carbon vacancies (V) as schematically depicted in FIG. \ref{fig:Figure1} a). An atomic force microscope scan of the ND containing the investigated SiV$^-$ revealed an agglomeration of NDs as shown in FIG. \ref{fig:Figure1} b). The dimensions of the cluster are of a small enough size to be integratable into a cavity system \citep{bayer2023}, while individual NDs that form the cluster are of a size where a modified phonon-density of states (PDOS) can be expected \citep{Klotz2022}. 
The electronic level scheme of the SiV$^-$ consists of four spin-degenerate orbital states, two of which form the ground-state (GS) and excited-state (ES), respectively. As a consequence four optically active transitions arise, which we label as A, B, C and D. For the remainder of the text we only use transition C, for which the spin levels are depicted in FIG. \ref{fig:Figure1} c). The spin degeneracy of the GS and ES levels can be lifted by applying a magnetic field, giving access to an electron spin qubit, e.g the one labeled by $\ket{\downarrow}$ and $\ket{\uparrow}$ \citep{Hepp2014, Rogers_electronic2014}.  When using such a spin-qubit at liquid helium temperature, its coherence time is mainly limited through phonon-induced dephasing. The latter can be mitigated by either cooling the system to mK temperatures \citep{Sukachev2017}, changing the PDOS \citep{Klotz2022} or increasing the GS-splitting, which suppresses phonon absorption \citep{Meesala2018, Sohn2018}. The use of NDs is an appealing choice as a host for the SiV$^-$, since they can combine two of the mentioned effects to increase spin-coherence times at temperatures around 4K. The reduced size modifies the PDOS and commonly-present strain in NDs results in an increased GS-splitting. We therefore investigate spectrally shifted SiV$^-$ where high strain can be expected \citep{Meesala2018} and  studied them for their optical and spin-coherence properties. \\
	%\textit{Results} -- 
The NDs were coated onto a sapphire substrate with good thermal conductivity. The sample was then cooled to liquid helium temperatures in a continuous flow-cryostat and investigated using a home-built confocal microscope. Four permanent magnets in a Hallbach-configuration designed for an in-plane field strength of around 400 mT were used to lift the spin degeneracy. Individual transitions of the SiV$^-$ were addressed by photo-luminescence excitation spectroscopy (PLE).
	\begin{figure}[t]
		\centering	
		\includegraphics{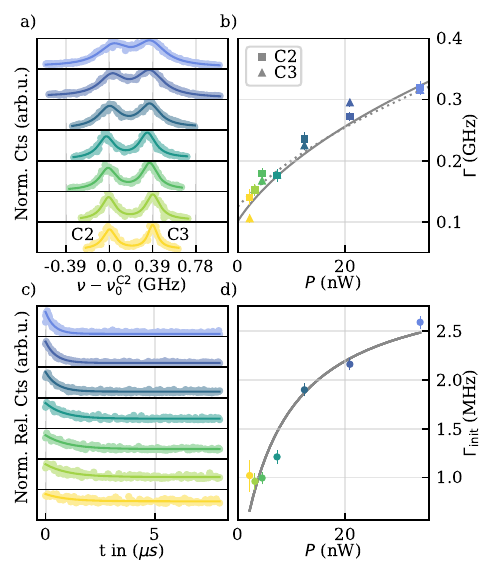}
		\caption{
			a) The data points show power-dependent PLE scans of the spin-preserving transitions C2 and C3 with the frequency relative to $\nu_0^{C_2}$ of transition C2. The solid lines are double-lorentzian fits to the data and reveal a splitting of $\protect\varr{spinsplittingr}\, \pm \protect\varr{errspinsplittingr}\SI{}{\mega\hertz}$ for the lowest excitation power $P$. 
			b) Fitted linewidth as a function of $P$ with squares and triangles representing C2 and C3, respectively. The dotted and solid line are the respective square-root fit. 
			c) Normalized fluorescence during spin initialization for increasing excitation powers with the solid lines being an exponential fit. For the highest power (lower panel) the initialization time is $\protect\varr{maxinitrater} \pm \protect\varr{errmaxinitrater} \SI{}{\nano\second}$ with a fidelity of $\protect\varr{maxfidelityr}\, \pm \, \protect\varr{errmaxfidelityr}$ . 
			d) The extracted initialization rates $\Gamma_\mathrm{init}$ as a function of $P$.}
		\label{fig:Figure2}
	\end{figure}
After finding a suitable SiV$^-$ the optical linewidths of the two spin-preserving transitions, labeled as C2 and C3, were investigated by PLE with varying power ($P$), as shown in FIG. \ref{fig:Figure2} a). The frequency splitting between C2 and C3 \varr{spinsplittingr}$\, \pm \,$\varr{errspinsplittingr} $\SI{}{\mega\hertz}$ was determined by a double-Lorentzian fit. FIG. \ref{fig:Figure2} b) shows the fitted power-dependent linewidth $\Gamma$ of C2 and C3. We extrapolated the linewidth to zero power using $\Gamma(P) = \Gamma_0 \sqrt{s + 1}$, where $\Gamma_0$ is the linewidth at zero power and $s=P/P_{\mathrm{sat}}$ with the saturation power $P_{\mathrm{sat}}$. The fit resulted in $\Gamma_0 ^{\text{C2}} = \varr{GammaC2r}\, \pm \,\varr{errGammaC2r}\, \SI{}{\mega\hertz}$  and $\Gamma_0 ^{\text{C3}} = \varr{GammaC3r}\, \pm \,\varr{errGammaC3r}\,\SI{}{\mega\hertz}$, respectively. The saturation powers turned out as $P_{\mathrm{sat}}^{\text{C2}} =\varr{psatC2r}\, \pm \,\varr{errpsatC2r}\SI{}{\nano\watt}$ and $P_{\mathrm{sat}}^{\text{C3}} =\varr{psatC3r}\, \pm \,\varr{errpsatC3r}\, \SI{}{\nano\watt}$, respectively. Firstly, $\Gamma_0$ is close to Fourier limits for commonly found lifetimes inside NDs, suggesting excellent optical quality. Secondly, the narrow linewidths compared to \varr{spinsplittingr} $\SI{}{\mega\hertz}$ allow for a strong drive without significant cross-talk between C2 and C3.\\ 
We proceeded by measuring the initialization rate of the spin for each excitation power. To this end, we pumped the population out of a thermal maximally-mixed equilibrium state with a resonant pulse of varying power. The resulting fluorescence, shown in FIG. \ref{fig:Figure2} c), was then fitted with an exponential decay. From the fit, we evaluated the initialization fidelity by comparing the maximum fluorescence at the beginning of the pulse with the fluorescence at the end of the pulse, corresponding to the steady state population under resonant drive.
%From the fit, we evaluated the initialization fidelity by comparing the maximum fluorescence to the one during the eventual steady state. 
In addition, the corresponding initialization rates $\Gamma_{\mathrm{init}}$ are shown in FIG. \ref{fig:Figure2} d). For the highest power we measured an initialization time of $\varr{maxinitrater}\, \pm \,\varr{errmaxinitrater}\SI{}{\nano\second}$ and a fidelity of $\varr{maxfidelityr}\, \pm \,\varr{errmaxfidelityr}$.
From the power dependence of $\Gamma_{\mathrm{init}}$, we determined the spin-branching ratio $\eta = \varr{etar}\, \pm \,\varr{erretar}$ by fitting $\Gamma_{\mathrm{init}} =\frac{1}{\eta}\frac{s}{(1+s)}\frac{\Gamma_0^{\text{C2}}}{2}$, where $\Gamma_0^{\text{C2}}$ and $P_{\mathrm{sat}}^{\text{C2}}$ are taken from the previously fitted PLE measurements \citep{Debroux2021}.\\
%The frequency splitting between C2 and C3 of (390$\pm$3) MHz in combination with the Zeeman-splitting of 10.956 GHz give insights on the possible GS-splitting of the SiV$^-$ \citep{Meesala2018}, indicating a high strain.
	\begin{figure*}[t]
		\centering
		\includegraphics{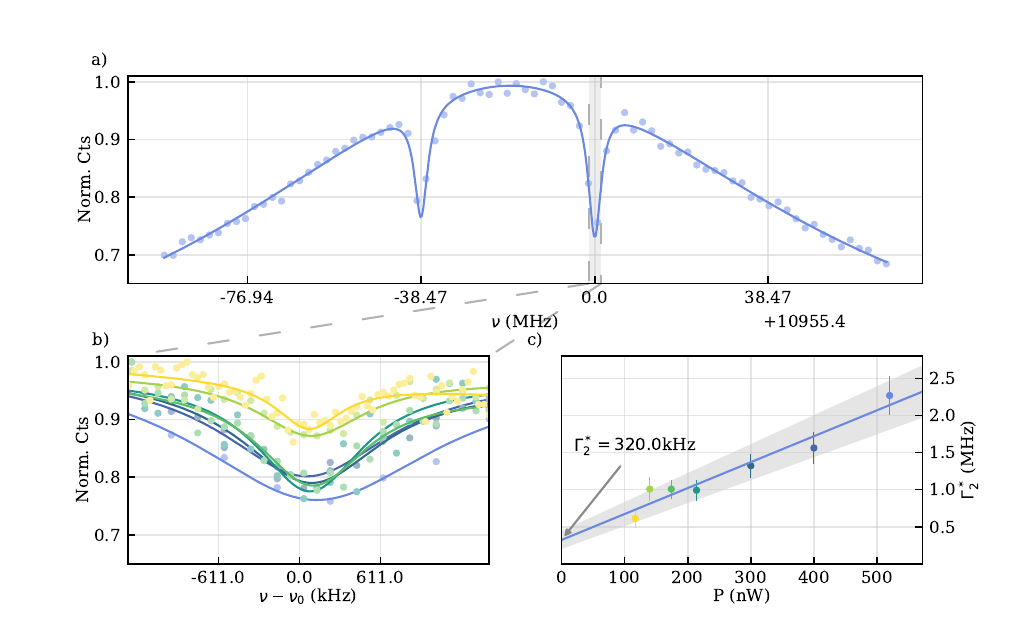}
		\caption{a) CPT measurement of a SiV$^-$-center showing two dips, indicating a nearby $^{13}C$ nuclear spin. The solid line is a triple-Lorentzian fit to the data points. The resulting Zeeman splitting is around 11GHz and the splitting between the two dips is $A = \protect\varr{Afig3r}\, \pm \, \protect\varr{errAfig3r} \, \SI{}{\mega\hertz}$. b) Measurement of the CPT width of the dip highlighted by the gray area in a) for varying power. c) Power dependent CPT width $\Gamma^\ast_2$. A linear fit yields a rate of $\Gamma^\ast_2 = \protect\varr{Gamma20r}\, \pm \, \protect\varr{errGamma20r} \, \SI{}{\kilo\hertz}$ at zero power.}
		\label{fig:Figure3}
	\end{figure*}
Extending the measurement procedure of FIG. 2c) to multiple consecutive resonant pulses with increasing temporal spacing probes the exponentially recovering spin population $T_1$. Fitting an exponential to the peak intensities in each fluorescence pulse resulted in $T_1 = \varr{T1r} \pm \varr{errT1r} \SI{}{\mu\second}$, which ultimately limits spin coherence. In addition, $T_1$ gives insights on the orientation of the SiV$^-$ symmetry axis to the magnetic field, where a misalignment leads to spin-mixing and hence shorter relaxation times. From the relatively short $T_1$ together with the spin-branching $\eta$ we estimate only a moderate alignment.\\
	% on the order of 10$^\circ$.\\
To probe the spin-coherence we used coherent population trapping (CPT). Here, the laser resonantly ($\Delta = 0$, see FIG.\ref{fig:Figure1} c)) drove a spin-flipping transition. Simultaneously, an electro-optical modulator (EOM) generated sidebands from which one was swept over the corresponding spin-preserving transition. If the Raman condition ($\delta = 0$) is fulfilled the system is pumped into a dark state quenching the fluorescence signal. For the studied SiV$^-$ two dips with a frequency splitting of $A = \varr{Afig3r}\, \pm \, \varr{errAfig3r} \, \SI{}{\mega\hertz}$ were present, as shown in FIG.\ref{fig:Figure3} a). The splitting $A$ is composed of two terms, the parallel coupling term $A_\parallel$ and the perpendicular coupling term $A_\perp$. The observed splitting is in close agreement with a theoretically predicted strongly hyperfine-coupled next-nearest neighbor $^{13}$C nuclear spin with a coupling strength of $A_\parallel = \SI{37}{\mega\hertz}$ \citep{nizovtsev2020}. \\ 
	%See SM for further discussion on hyperfine coupling under various system parameters.\\ 
Furthermore, since the dip's linewidth gives insight on the spin's dephasing rate, we performed a power dependent measurement, which is shown in FIG. \ref{fig:Figure3} b). Fitting the data with a Lorentzian and extracting the linewidth  $\Gamma_2^\ast$ allowed us to linearly extrapolate the dephasing rate to zero laser power to suppress power-induced broadening. As a result, we obtained a zero-power linewidth of $\Gamma^\ast_2 = \varr{Gamma20r}\, \pm \, \varr{errGamma20r} \, \SI{}{\kilo\hertz}$. 
This is approximately a ten-fold reduced decoherence rate compared to previously reported measurements in bulk diamond at similar temperatures with $\Gamma^\ast_2 \approx \SI{4.5}{\mega\hertz}$ \citep{Rogers2014_Spin, Becker2018, Pingault2014}, and a minor improvement to measurements in a strain-engineered nano-beam, which reported $\Gamma^\ast_2 = \SI{640}{\kilo\hertz}$ \citep{Sohn2018, Meesala2018}. 
Combining the latter improvements of spin coherence in high strain environments with the present spin transition frequency of roughly \SI{11}{\giga\hertz} and the spin-preserving transitions' splitting of \SI{390}{\mega\hertz} suggests a comparably highly strained SiV$^-$-center with a GS splitting on the order of 500 GHz \citep{Meesala2018}. \\
	%See SM for discussion on spin splittings for various GS and field configurations.\\
	\begin{figure*}[t]
		\centering
		\includegraphics{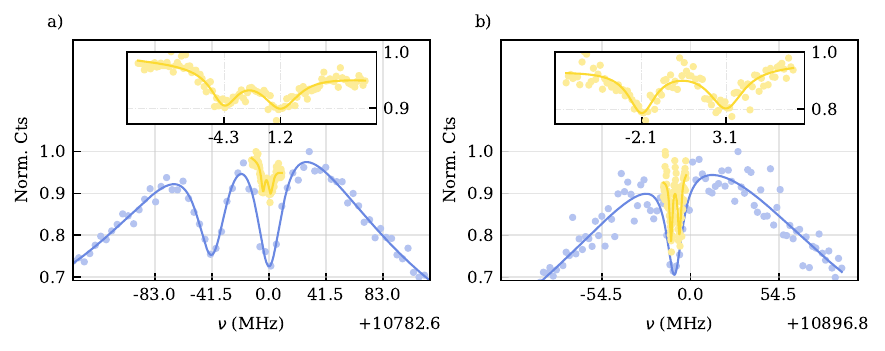}
		\caption{CPT measurements of two more SiV$^-$ with low and high optical power, indicated in yellow and blue, respectively. a) shows two dips, split by  $A = \protect\varr{Afig4ar}\, \pm \, \protect\varr{errAfig4ar} \, \SI{}{\mega\hertz}$. Low-power measurements reveal a coupled system with an additional splitting of $A' = \protect\varr{Apfig4ar}\, \pm \, \protect\varr{errApfig4ar} \, \SI{}{\mega\hertz}$. b) Measurements shows a double dip, split by $5.20 \pm 0.14$ MHz, indicating a weakly coupled C13 nuclear spin.}
		\label{fig:Figure4}
	\end{figure*}
We further investigated the spin environment in the same agglomeration of NDs by performing CPT measurements on different SiV$^-$-centers. The measurement revealed coupling of multiple spins, as shown in FIG. \ref{fig:Figure4}. For example, the SiV$^-$ in FIG \ref{fig:Figure4} a) displayed two dips split by  $A = \varr{Afig4ar}\, \pm \, \varr{errAfig4ar} \, \SI{}{\mega\hertz}$. Measurements with reduced power resolved two more dips, split by  $A' = \varr{Apfig4ar}\, \pm \, \varr{errApfig4ar} \, \SI{}{\mega\hertz}$ as shown in the inset. They exhibited linewidths of $\varr{Gammafig4alr} \pm \varr{errGammafig4alr} \SI{}{\mega\hertz}$ and $\varr{Gammafig4arr} \pm \varr{errGammafig4arr} \SI{}{\mega\hertz}$ , for the left and right dip, respectively. The distinct linewidths suggest coupling to two other surrounding SiV$^-$-centers' electron spins with coherence properties commonly found with SiV$^-$ in low to moderate strain bulk diamond \citep{Rogers2014_Spin, Becker2018, Pingault2014}. Assuming a dipolar electron-electron coupling would correspond to a distance on the order of 10 \SI{}{\nano\meter}, reasonable for the size of the ND and density of SiV$^-$-centers under study. \\
Another SiV$^-$, displayed in FIG. \ref{fig:Figure4} b), exhibited two individual dips at low optical power with linewidths of $\varr{Gammafig4blr} \pm \varr{errGammafig4blr} \SI{}{\mega\hertz}$  and $\varr{Gammafig4brr} \pm \varr{errGammafig4brr} \SI{}{\mega\hertz}$, for the left and right respectively. In contrast to the previous SiV$^-$s' multi-dip structure with distinct linewidths, the fact that the present SiV$^-$ has closely matching and relatively narrow linewidths is indicative of a coupled nearby $^{13}$C nuclear spin with a coupling strength of $5.20 \pm 0.14$MHz \citep{Sohn2018}. In this case the distance between the two coupled spins is on the order of $\SI{1}{\angstrom}$ \citep{Sohn2018}.
	
	%\textit{Outlook} -- 
The presented results open up new possibilities to utilize the  electron and nuclear spin environment of SiV$^-$-centers in NDs as an elementary unit of diamond-based qubits with integrability into photonic platforms, like photonic crystal or open microcavities \citep{antoniuk2023alloptical, bayer2023, Nguyen2019, Fehler2021}. The thereby formed hybrid quantum system enables efficient mapping of the quantum state of the electron spin to flying photonic qubits as well as coupling to local memory units consisting of nearby nuclear spins. Our work also suggests that the access to small spin registers is feasible for SiV$^-$-centers in NDs.
	%Furthermore, going to even higher strain might push the dephasing times of the SiV$^-$ to the $\mu$s regime at temperatures above 4 K. 
Increasing the GS-splitting beyond 500 GHz, e.g. by hydrostatic pressure \citep{vindolet2022} might further improve coherence times. Additionally temperatures around 4 K within a flow-cryostat become sufficient. The measured dephasing rates bring coherent control of different spin qubits in NDs, either by means of direct microwave drive \citep{Pingault2017} or all-optical control using a Raman-type lambda scheme, as has recently been shown for the tin-vacancy \citep{Debroux2021}, into reach.  For the system parameters in our measurements, an all-optical Rabi driving strength in the order of several MHz can be expected, comparable to driving rates achieved with microwave driving \citep{Pingault2017} . Additionally, the high sensitivity enables to detect close-by electron or nuclear spins. Furthermore, with coherent electron spin control, direct \citep{Stas2022} or indirect \citep{Laucht2022} control of a nuclear spin becomes possible. Mapping information of the electron spin to a strongly coupled long-lived nuclear spin establishes SiV$^-$-centers in NDs as a viable candidate for an interchangeable hybrid quantum memory. 
	
\begin{acknowledgments}
		The authors thank V.A. Davydov for synthesis and processing of the nanodiamond material. The project was funded by the Baden-Württemberg Stiftung in Project Internationale Spitzenforschung. N.L acknowledges support from IQST. A.K. acknowledges support of the BMBF/VDI in the Projects HybridQToken (16KISQO43K), QR.X (16KISQ006) and Spinning (13N16215). The authors acknowledge funding by the European Union and the DFG within the Quantera-project SensExtreme.
	\end{acknowledgments}

\end{document}